\documentclass[aps,prl,twocolumn,amsmath,amssymb,showpacs, superscriptaddress,notitlepage,longbibliography]{revtex4-2}

\usepackage[colorlinks=true,linkcolor=blue,anchorcolor=red,citecolor=blue, urlcolor=blue]{hyperref}
\usepackage{bm}
\usepackage{graphicx}
\usepackage{color}
\usepackage{soul}

\begin{document}


\title{
Absence of the anomalous Hall effect in planar Hall experiments}

\author{C. M. Wang}
\affiliation{Department of Physics, Shanghai Normal University, Shanghai 200234, China}
\affiliation{Quantum Science Center of Guangdong-Hong Kong-Macao Greater Bay Area (Guangdong), Shenzhen 518045, China}
\affiliation{Shenzhen Key Laboratory of Quantum Science and Engineering, Shenzhen 518055, China}
\affiliation{Shenzhen Institute for Quantum Science and Engineering and Department of Physics, Southern University of Science and Technology (SUSTech), Shenzhen 518055, China}
\affiliation{International Quantum Academy, Shenzhen 518048, China}

\author{Z. Z. Du}
\affiliation{Shenzhen Institute for Quantum Science and Engineering and Department of Physics, Southern University of Science and Technology (SUSTech), Shenzhen 518055, China}
\affiliation{Quantum Science Center of Guangdong-Hong Kong-Macao Greater Bay Area (Guangdong), Shenzhen 518045, China}
\affiliation{Shenzhen Key Laboratory of Quantum Science and Engineering, Shenzhen 518055, China}
\affiliation{International Quantum Academy, Shenzhen 518048, China}

\author{Hai-Zhou Lu}
\email{Corresponding author: luhz@sustech.edu.cn}
\affiliation{Shenzhen Institute for Quantum Science and Engineering and Department of Physics, Southern University of Science and Technology (SUSTech), Shenzhen 518055, China}
\affiliation{Quantum Science Center of Guangdong-Hong Kong-Macao Greater Bay Area (Guangdong), Shenzhen 518045, China}
\affiliation{Shenzhen Key Laboratory of Quantum Science and Engineering, Shenzhen 518055, China}
\affiliation{International Quantum Academy, Shenzhen 518048, China}

\author{X. C. Xie}
\affiliation{International Center for Quantum Materials, School of Physics, Peking University, Beijing 100871, China}
\affiliation{Institute for Nanoelectronic Devices and Quantum Computing, Fudan University, Shanghai 200433, China}
\affiliation{Hefei National Laboratory, Hefei 230088, China}

\begin{abstract}
Recently, the planar Hall effect has attracted tremendous interest. In particular, an in-plane magnetization can induce an anomalous planar Hall effect with a $2\pi/3$ period for hexagon-warped energy bands. This effect is similar to the anomalous Hall effect resulting from an out-of-plane magnetization. However, this anomalous planar Hall effect is absent in the planar Hall experiments. Here, we explain its absence, by performing a calculation that includes not only the Berry curvature mechanism as those in the previous theories, but also the disorder contributions. {The conventional $\pi$-period planar Hall effect will occur if the mirror reflection symmetry is broken, which buries the anomalous one.} We show that an in-plane strain can enhance the anomalous Hall conductivity and changes the period from $2\pi/3$ to $2\pi$. We propose a scheme to extract the hidden anomalous planar Hall conductivity from the experimental data. Our work will be helpful in detecting the anomalous planar Hall effect and could be generalized to understand mechanisms of the planar Hall effects in a wide range of materials. 
\end{abstract}

\date{\today}

\maketitle

{\color{blue}\emph{Introduction.}} -- In the planar Hall effect, an in-plane transverse voltage can be induced by coplanar electric and magnetic fields, irrelevant to the Lorentz force. It used to be expected only in ferromagnetic and antiferromagnetic materials \cite{Ky1968pssb,Tang2003prl,Shin2007prb,Seemann2011prl,Cao2011apl,Yin2019prl}, but recently, it has attracted much attention in nonmagnetic materials. So far, there are three mechanisms. (i) the conventional mechanism \cite{taskin2017nc,Huang2021prr,Zheng2020prb,Rao2021prb,Yamada2021arXiv};
(ii) the chiral anomaly \cite{Burkov2017prb,Nandy2017prl,Nandy2018sr,Imran2018prb,Ma2019prb,Kundu2020njp,Sharma2020prb,Ahmad2021prb,Ge2020nsr}, which is excluded in 2D as chirality is defined in odd dimensions; and (iii) the newly proposed mechanism, the anomalous planar Hall effect induced by an in-plane magnetization, 
such as in spin-orbit coupled 2D systems \cite{Zhang2011prb,Zyuzin2020prb,Cullen2021prl,Battilomo2021prr}, atomic crystals \cite{Ren2016prb,Zhong2017prb}, topological materials \cite{Liu2013prl,Akzyanov2018prb,Liu2018prl,Guo2020nl,Tan2021prb,Jin2021ns,Sun2022prbl,Wang2022arxiv}, and heterodimensional superlattice \cite{Zhou2022nature,Cao2023prl}. 
In the third mechanism, in-plane magnetization can be tuned by an in-plane external magnetic field without introducing the Lorentz force.

However, the anomalous planar Hall effect is still  absent in the planar Hall measurements. For example, it has been theoretically suggested that the hexagonal warping in the band structure, such as on the surface of a topological insulator \cite{Liu2013prl,Akzyanov2018prb,Sun2022prbl}, could lead to a $2\pi/3$ period of the anomalous planar Hall effect. However, measurements in the topological insulators Bi$_{2-x}$Sb$_{x}$Te$_{3}$ \cite{taskin2017nc}, Sn-doped Bi$_{1.1}$Sb$_{0.9}$Te$_{2}$S \cite{Wu2018apl}, and Bi$_2$Te$_3$ \cite{Bhardwaj2021apl} all show only a $\pi$ period, despite the observed strong warping of these materials \cite{Fu2009prl,Hasan2009physics,Zhanybek2010prl,Kuroda2010prl}. 

\begin{figure}[thbp]
	\centering
	\includegraphics[width=\columnwidth]{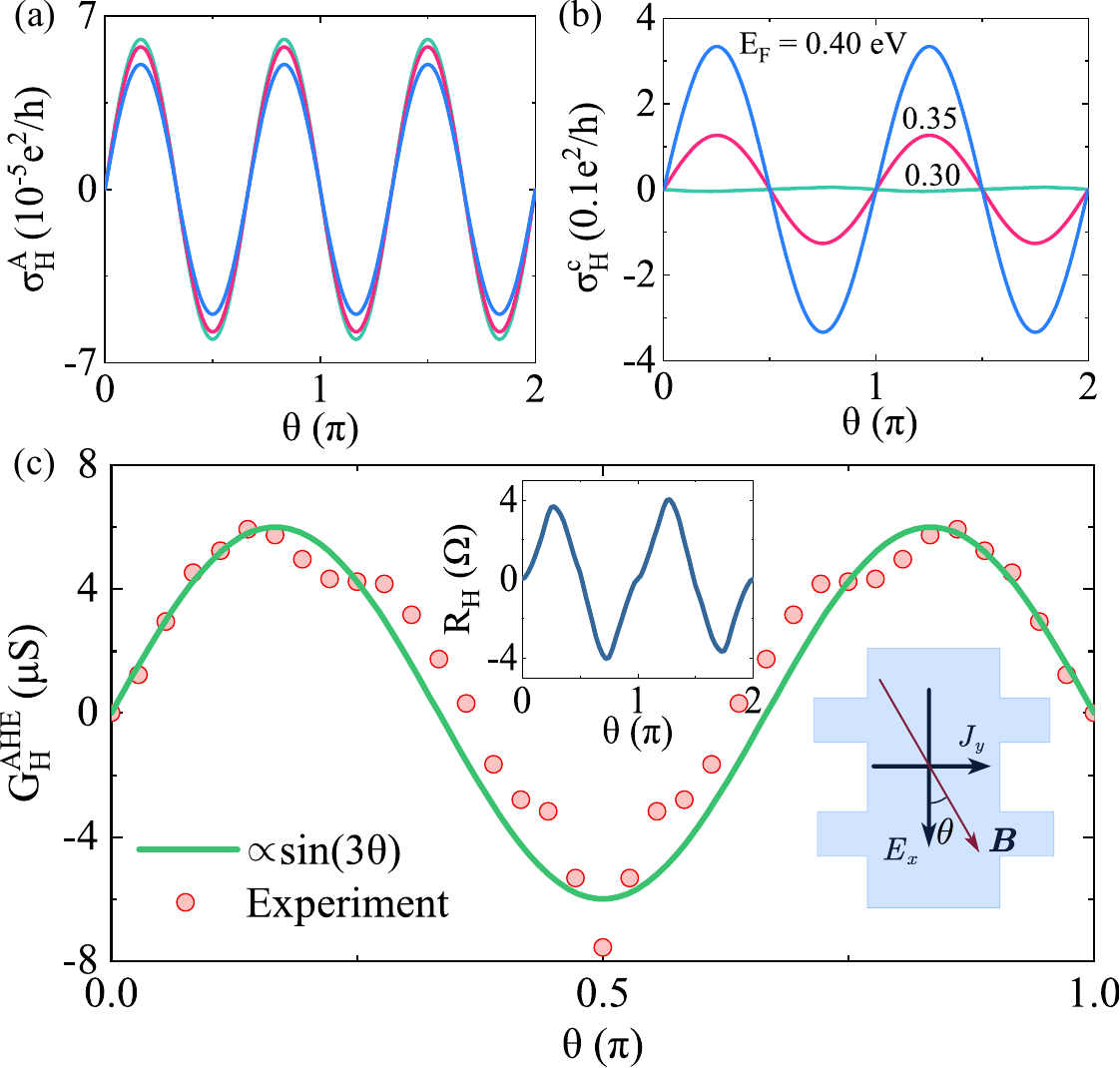}
	\caption{[(a) and (b)] Comparison between the theoretically calculated anomalous planar Hall conductivity $\sigma_\mathrm{H}^\mathrm{A}$ with a $2\pi/3$ period and conventional planar Hall conductivity $\sigma_\mathrm{H}^\mathrm{c}$ with the $\pi$ period. The parameters are $v=0.3$ eVnm, $\alpha=0.2$ eVnm$^3$, $\Delta=0.01$ eV, $\tau=2\hbar v^2/(n_iu_0^2E_F)=1$ ps,  and $r={(n_iu_1^3)^{2/3}}/({n_iu_0^2})=0.1$. (c) The anomalous Hall conductance converted from the measured Hall resistance $R_\text{H}$ (inset) and longitudinal resistance $R_{xx}$ in the experiment \cite{Wu2018apl}, as the difference (divided by 2) between the original Hall conductance data in the range of $\theta\in [0,2\pi]$ and its mirror reflection with respect to $\theta=\pi$.}\label{Fig:fig1}
\end{figure}

\begin{table}[htbp]
	\caption{Comparison between the anomalous and conventional planar Hall effects, in the presence of the hexagonal warping with 3-fold rotational symmetry and in-plane strain at a relaxation time $\tau$ of about $1$ ps. $B$ is the in-plane magnetic field. The period is defined by $\theta$ in Fig. \ref{Fig:fig1}. $e^2/h$ is the conductance quantum. {Here $\theta$ is the angle between the magnetic and electric fields.}
	}
	\label{Tab:Hall}
	\begin{ruledtabular}
		\begin{tabular}{ccc}
			& Anomalous & Conventional  \\
			Magnitude &  $10^{-5} e^2/h$   & $10^{-1} e^2/h$    \\
			$B$-dependence &  $B^3$   &  $B^2$\\
			Period (warping)&  $2\pi/3$  & $\pi$  \\
			Period (in-plane strain) &  $2\pi$  & $\pi$  \\
		\end{tabular}
	\end{ruledtabular}
\end{table}

In this Letter, we propose a theory to understand the absence of the anomalous planar Hall effect in the experiments. The anomalous planar Hall effect requires breaking of all mirror-reflection symmetries \cite{Liu2013prl} in addition to time-reversal symmetry breaking. However, the breaking of the mirror-reflection symmetry may also induce the conventional planar Hall effect with considerably large values because it is inversely proportional to the impurity density in diffusive systems. As a result, {the total planar Hall conductivity is nearly identical to the conventional one, thereby implying that} the anomalous planar Hall effect is covered by the conventional planar Hall effect. To illustrate our explanation, we use the hexagon-warped (threefold rotational symmetric) 2D surfaces states in a topological insulator as an example and calculate the planar Hall conductivity in the presence of an in-plane magnetic field. {We further extend the previous researches \cite{Liu2013prl,Akzyanov2018prb,Sun2022prbl} on the anomalous planar Hall effect by including disorder in the analysis.}
Up to the first order of the warping, only the anomalous planar Hall effect contributes to the Hall conductivity. Both the intrinsic and non-crossing impurity-related extrinsic Hall conductivities are independent of the impurity density, except for the small higher-order skew scattering contribution. They exhibit a $2\pi/3$ period (Fig. \ref{Fig:fig1}(a)) and a $B^3$-dependence on the in-plane magnetic field,  Surprisingly, the conventional planar Hall effect occurs beyond the first order of the warping, with a $B^2$ dependence and $\pi$ period. Further, it is stronger than the anomalous planar Hall effect by several orders of magnitude (Fig. \ref{Fig:fig1}(b)). Hence, the total planar Hall conductivity has a $\pi$ period, as those observed in the experiments. 
Because of the huge difference in magnitude, it is difficult to use the Fourier transformation to extract the anomalous planar Hall conductance from the experimental data. We propose to extract the anomalous planar Hall conductance as the difference (divided by 2) between two planar Hall conductance data sets. They are obtained by measuring the conductance as a function of the magnetic field angle $\theta\in [0,2\pi]$ and its mirror reflection with respect to $\theta=\pi$. We apply this data analysis scheme to the experiment of the topological insulator Sn-doped Bi$_{1.1}$Sb$_{0.9}$Te$_2$S \cite{Wu2018apl}. It shows the extracted anomalous planar Hall conductance follows a $\sin(3\theta)$ dependence on the magnetic field angle with a $2\pi/3$ period (Fig. \ref{Fig:fig1}(c)). Our work will be helpful for further investigations on the mechanisms of the planar Hall effects.

\begin{figure}[thbp]
	\centering
	\includegraphics[width=0.9\columnwidth]{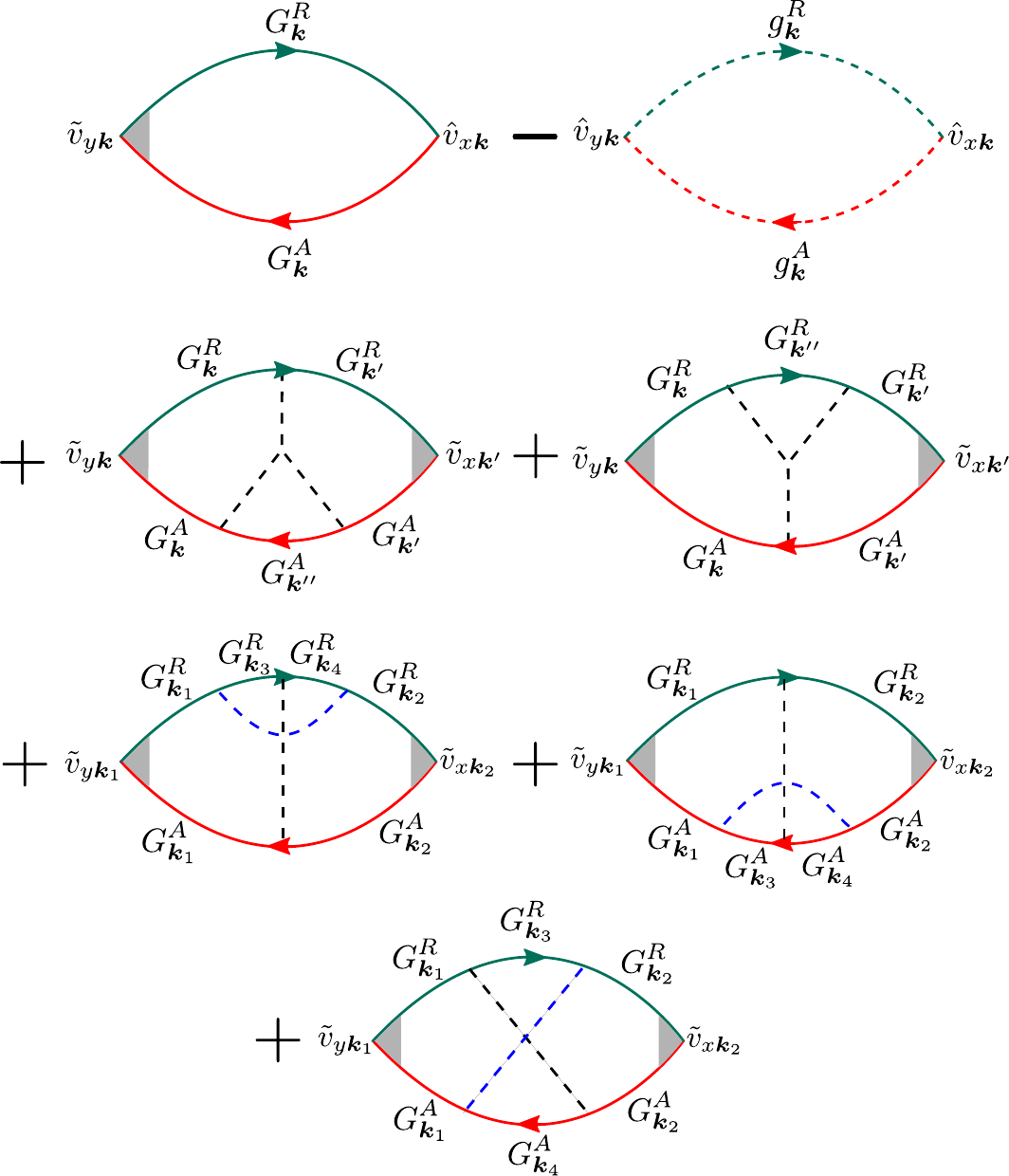}
	\caption{The Feynman diagrams for the extrinsic contribution to the planar Hall conductivity. The arrowed solid and dashed lines stand for the full and bare Green's functions $ G_{\bm k}^{R/A}$  and $g_{\bm k}^{R/A}$, respectively. 
	The gray shadows represent the vertex correction to the velocity from the ladder diagrams. The dashed lines without arrows represent the disorder scattering. We model disorder as  randomly located nonmagnetic $\delta$-function impurities with the strength distributions satisfying $\langle u_i\rangle_{\mathrm{imp}}=0$, $\langle u_i^2\rangle_{\mathrm{imp}}=u_0^2$, and $\langle u_i^3\rangle_{\mathrm{imp}}=u_1^3$. The velocity $\hat v_\mu$ ($\mu=x,y$) is corrected by disorder as
$\tilde v_{\mu}(\bm k)|_{k=k_F(\phi)}
	=\hat v_{\mu}(\bm k)|_{k=k_F(\phi)} 
	+n_iu_0^2\sum_{\bm k'} G_{\bm k'}^A(E_F)\tilde v_{\mu}(\bm k') G_{\bm k'}^R(E_F)$, where
$n_i$ is the impurity density, and $k_F(\phi)$ is the Fermi wave vector with $\phi$ being the polar angle. }\label{Fig:surface}
\end{figure}

{\color{blue}\emph{Disorder-corrected planar Hall conductivity.}} 
-- Beyond the previous works \cite{Liu2013prl,Akzyanov2018prb,Sun2022prbl} in which the Berry curvature is the only mechanism (intrinsic part), we also take into account the disorder contributions (extrinsic part) in the calculation of the anomalous planar Hall conductivity \cite{Supp-Weyl}. The intrinsic part is given by the summation of the $z$-component Berry curvature $\Omega^\lambda_{z\bm k}$ of the occupied states  
\begin{align}\label{eq:Fermi-sea}
	\sigma_\text{H}^\text{in}=\frac{e^2}{\hbar}\sum_{\lambda\bm k}\Omega_{z\bm k}^\lambda n_\text{F}(E_{\lambda\bm k}),
\end{align}
where $n_\text{F}$ is the Fermi distribution at the $\lambda$th band $E_{\lambda\bm k}$. 
The extrinsic part from the disorder scattering can be calculated \cite{Supp-Weyl} using the Feynman diagrams in Fig. \ref{Fig:surface}

We consider the hexagon-warped surface states of a topological insulator, in an in-plane magnetic field $\bm B$
\begin{align}
	\hat H=v\left(k_{x} {\sigma}_{y}-k_{y} {\sigma}_{x}\right)+\frac{\alpha}{2}\left(k_{+}^{3}+k_{-}^{3}\right) {\sigma}_{z}+\Delta_x\sigma_x+\Delta_y\sigma_y,
\end{align}
where $v$ is the Dirac velocity, $\alpha$ measures the hexagonal warping that brings the three-fold rotational symmetry, the in-plane magnetic field comes in terms of the Zeeman splitting $\bm \Delta=(\Delta_x,\Delta_y)=\Delta(\cos\theta,\sin\theta)=g\mu_B\bm B/2$ with $g$ being the effective g-factor, $\bm \sigma=(\sigma_x,\sigma_y,\sigma_z)$ are the Pauli matrices, and $k_{\pm}=k_{x} \pm i k_{y}$. 
{The orbital contribution from the in-plane magnetic field relates to the $z$-component of the vector potential, which will not enter into the two-dimensional Hamiltonian due to the Peierls substitution. Hence, there is no orbital contribution excluding the Lorentz force,} thus the planar Hall effect is purely induced by the in-plane magnetization due to the in-plane magnetic field. The warping term is the key to realize the anomalous planar Hall effect since the $\sigma_z$ term is essential for the nonzero $z$-component Berry curvature. The warping coefficient could reach a considerable value up to $0.25$ eVnm$^3$ in the surface states of Bi$_2$Te$_3$-type topological insulators \cite{Fu2009prl,Zhanybek2010prl}. This model carries
$\Omega^\lambda_{z \bm k}=\frac{\lambda v\alpha}{2\varepsilon_{\bm k}^3} [2v k_x(k_x^2-3k_y^2)+3(\Delta_yk_x^2+2\Delta_xk_xk_y-\Delta_yk_y^2) ]$ with $\lambda=\pm1$. The expressions of eigenenergies including $\varepsilon_{\bm k}$ and velocities can be found in \cite{Supp-Weyl}.

Firstly, we study the intrinsic and non-crossing extrinsic contributions. Up to the first order of $\alpha$, we find the anomalous planar Hall conductivity $\sigma_{\text{H}}\equiv\sigma_\text{H}^\text{in}+\sigma_\text{H}^\text{ex}$ as \cite{Supp-Weyl}
\begin{align}
	\sigma_{\text{H}} =\frac{e^2}{h}\left(\frac{4}{E_F}   + \frac{u_1^3}{n_iu_0^4}\right)\frac{\alpha}{v^3}\Delta^3\sin3\theta ,\label{total_ahe}
\end{align}
where the third-order skew-scattering disorder correlations (the third and fourth diagrams in Fig. \ref{Fig:surface}) results in the second term that is inversely proportional to the impurity density $n_i$. On the other hand, the intrinsic (Berry curvature) as well as the side-jump and other skew-scattering disorder contributions give rise to the first term that has nothing to do with $n_i$.
Both terms $\propto B^3$ and shows three-fold rotational symmetry.
The $n_i$-independent part is eight times of the intrinsic (i.e., Berry curvature alone) planar Hall conductivity \cite{Sun2022prbl,Supp-Weyl}.
Up to the first order of $\alpha$, the leading disorder contributions are fully due to the anomalous planar Hall effect and could strongly influence the intrinsic contribution. The term in the parenthesis could be rewritten as $4/E_F+r^{1.5}\sqrt{E_F\tau/(2\hbar v^2)}$ with $r$ being defined in Fig. \ref{Fig:fig1}. The parameter $r$ determines the relative strength of the third-order skew-scattering contribution compared to other contributions.

\begin{figure}[thbp]
	\centering
    \includegraphics[width=\columnwidth]{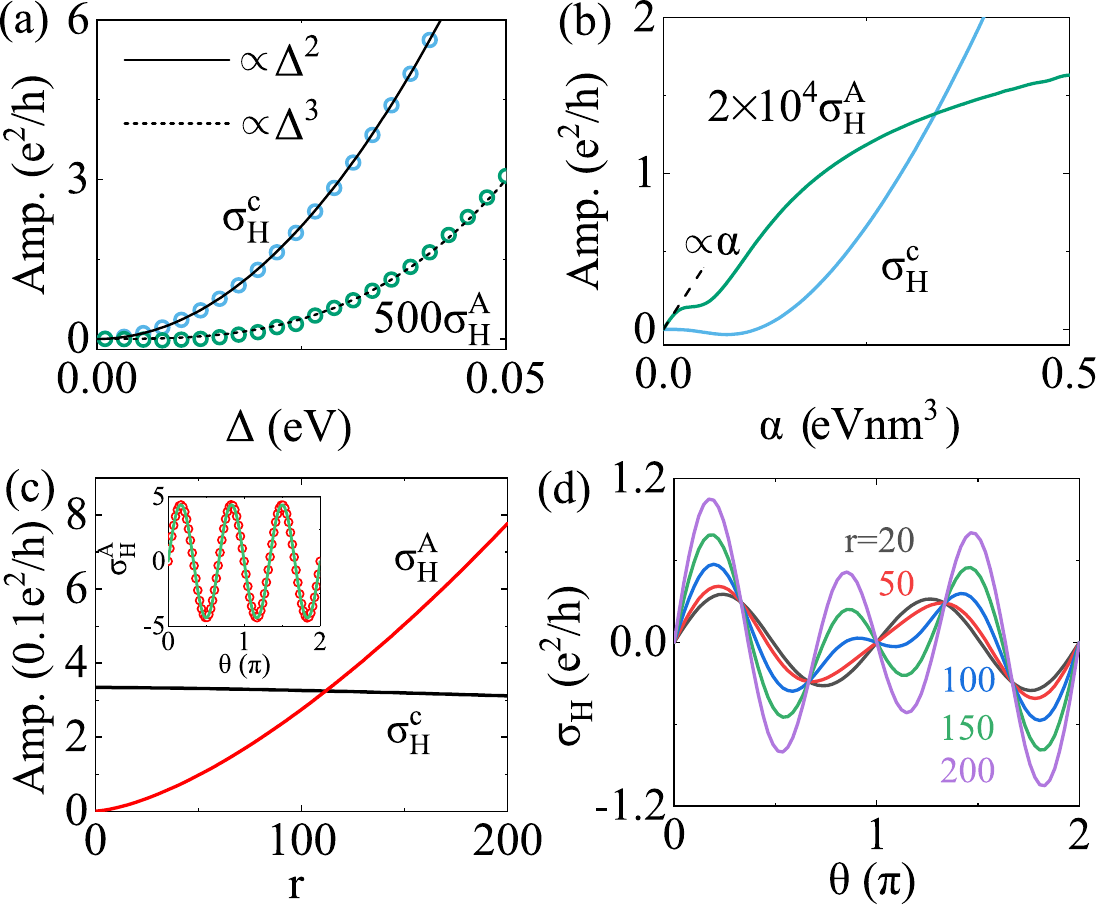}
	\caption{The amplitudes of the anomalous and conventional Hall conductivities as functions of the Zeeman energy shown with circles (a), the warping coefficient $\alpha$ (b), and the dimensionless ratio $r={(n_iu_1^3)^{2/3}}/{n_iu_0^2}$ that describes the third-order skew scattering (c). The inset of (c) shows the anomalous Hall conductivity for the weak warping coefficient $\alpha=0.001$ eVnm$^3$, where the circles are the numerical results while the green line is from Eq. \eqref{total_ahe}. (d) The total Hall conductivities $\sigma_\text{H}$ of different $r$ are plotted as functions of the angle $\theta$. The Fermi energy $E_F=0.4$ eV, the Fermi velocity $v=0.3$ eVnm and $\tau=1$ ps. In (a) and (b), $r=0.1$. In (b)-(d), $\Delta=0.01$ eV. In (a), (c), and (d), $\alpha=0.2$ eVnm$^3$.}\label{Fig:fig2}
\end{figure}

{\color{blue}\emph{Why anomalous much weaker than conventional?}} -- Our theory can deal with the anomalous and conventional planar Hall effects in a unified way that fully takes into account the higher-order warping terms. We achieve this by numerical calculation with realistic sample parameters. The inset of Fig. \ref{Fig:fig2}(c) shows that the numerical $\sigma_{\text{H}}^\text{A}$ (the circles) at a weak warping $\alpha=0.001$ eVnm$^3$ agrees well with the analytical Eq. \eqref{total_ahe}. This comparison confirms the accuracy of the numerical calculation. Beyond the weak warping limit, the 
	mirror symmetry breaking will also results in a large conventional Hall conductivity inversely proportional to the impurity density. It is in vivid contrast to the out-of-plane magnetization case, where only the anomalous Hall effect exists \cite{Nagaosa2010rmp}.
The total planar Hall conductivity, which is the sum of $\sigma_\mathrm{H}^\mathrm{A}$ (anomalous) and $\sigma_\mathrm{H}^\mathrm{c}$ (conventional) in Figs. \ref{Fig:fig1}(a) and (b), exhibits approximately the same $\sin 2\theta$ behavior with the periodicity $\pi$ as $\sigma_\mathrm{H}^\mathrm{c}$ in Fig. \ref{Fig:fig1}(b). The conventional $\sigma_\text{H}^\text{c}$ can be approximated by the classical Boltzmann one $\sigma_\text{H}^\text{Boltz}=-e^2\tau \sum_{\bm k}v_xv_y\delta(\varepsilon_{\bm k}-E_F)$ with the group velocities $v_{x,y}=\hbar^{-1}\partial_{k_{x,y}}\varepsilon_{\bm k}$. Up to the lowest order of the warping and the magnetic field,
\begin{align}
	\sigma_\text{H}^\text{Boltz}=\frac{e^2}{h}\frac{E_F\tau}{h}\frac{27\pi E_F^2}{4v^6}\Delta^2\alpha^2\sin2\theta.\label{Boltz}
\end{align}	
In the weak scattering regime ($\tau\sim $ ps), the dimensionless quantity $E_F\tau/h$ is very large, which in turn results in a significant disparity between the anomalous $\sigma_{\text{H}}^\text{A}$ and conventional $\sigma_{\text{H}}^\text{c}$.
\begin{eqnarray}
	\sigma_{\text{H}}^\text{A}\sim \frac{e^2}{h}\frac{4\alpha  }{ v^3E_F} \Delta^3 \ll \sigma_{\text{H}}^\text{c}
	\sim \sigma_\text{H}^\text{Boltz}.
\end{eqnarray}
Here the weak third-order impurity scattering term is ignored in $\sigma_{\text{H}}^\text{A}$. In summary, beyond the first order of warping, the conventional planar Hall conductivity exhibits a significantly larger value that obscures the anomalous one.

We can distinguish the small anomalous planar Hall effect from the conventional planar Hall effect, according to their different magnetic field dependence, as shown in Figs. \ref{Fig:fig1}(a) and (b). As expected, $\sigma_{\text{H}}^\text{A}$ oscillates with a $2\pi/3$ period, while the period of $\sigma_{\text{H}}^\text{c}$ is $\pi$. This $\pi$ period originates from the conventional planar Hall part in $\sigma_\text{H}^\text{ex}$ and $\propto 1/n_i$ as indicated by Eq. \eqref{Boltz}. The conventional planar Hall conductivity
\begin{eqnarray}\label{sigma_c}
	\sigma_{\text{H}}^\text{c}(\bm B)=\sigma_{\text{H}}^\text{c}(-\bm B),
\end{eqnarray}
which has a $\sin2\theta$ dependence. By contrast, the anomalous  $\sigma^\mathrm{A}_\mathrm{H}\propto\sum_{\lambda\bm k}\Omega_z^\lambda$, where the Berry curvature is anti-symmetrical $\Omega_z^\lambda(-\bm B,-\bm k)=-\Omega_z^\lambda(\bm B,\bm k)$, then
\begin{eqnarray}\label{sigma_A}
	\sigma_{\text{H}}^\text{A}(\bm B)=-\sigma_{\text{H}}^\text{A}(-\bm B).
\end{eqnarray}
This means the anomalous part is an odd function of the in-plane magnetic field. The amplitudes of $\sigma_{\text{H}}^\text{A}$ and $\sigma_{\text{H}}^\text{c}$ versus the Zeeman energy in Fig. \ref{Fig:fig2}(a) show the approximate relations  $\sigma_{\text{H}}^\text{A}\propto \Delta^3$ and $\sigma_{\text{H}}^\text{c}\propto \Delta^2$, which verify the argument in Eqs. \eqref{sigma_c} and \eqref{sigma_A}.

{\color{blue}\emph{Dependence on Fermi energy and warping}} -- 
For a small $r$ (defined in Fig. \ref{Fig:fig1}), the amplitude of the anomalous planar Hall conductivity decreases with the increasing Fermi energy roughly following $1/E_F$ as shown in Fig. \ref{Fig:fig1}. This behavior is in agreement with Eq. \eqref{total_ahe}, and is in sharp contrast with the conventional planar Hall effect whose amplitude is enhanced dramatically for $E_F>0.3$ eV. Moreover, the amplitude of the conventional one does not vary with the increasing Fermi energy, monotonously. At the energy $E_F\sim0.30$ eV, the conductivity almost vanishes. This complicated behavior implies the subtlety of the planar Hall effects.

Figure \ref{Fig:fig2}(b) illustrates the amplitudes of $\sigma_{\text{H}}^\text{A}$ and $\sigma_{\text{H}}^\text{c}$ as functions of the warping coefficient $\alpha$. Both the anomalous and the conventional planar Hall effects vanish in the limit $\alpha\rightarrow 0$, indicating the important role of the warping in the planar Hall effect. The amplitude of the anomalous planar Hall conductivity increases monotonously with $\alpha$, obeying the linear relation at weak $\alpha$ in the analytical Eq. \eqref{total_ahe}, while tends to saturate in the large-warping limit. 
For small warping, $\sigma_\text{H}^\text{c}$ increases very slowly from zero, even shows a small and negative dip near $\alpha=0.1$ eVnm$^3$, so the total planar Hall conductivity Eq. \eqref{total_ahe} includes only the anomalous contribution up to the first order of $\alpha$. Nevertheless, the amplitude of $\sigma_\text{H}^\text{c}$ grows rapidly at large $\alpha$. It is worth noting that we do not assume any anisotropic scattering or tilted energy band, which differs from previous works \cite{taskin2017nc,Zheng2020prb,Rao2021prb}.

\begin{figure}[thbp]
	\centering
	\includegraphics[width=\columnwidth]{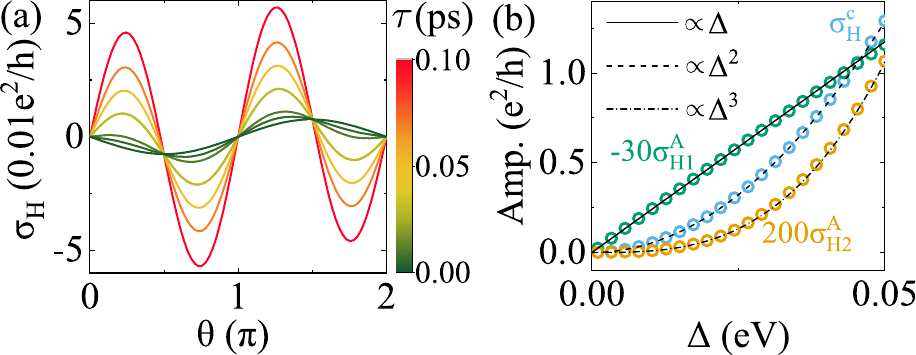}
	\caption{(a) In the presence of an in-plane strain, the planar Hall conductivity $\sigma_\text{H}$ at different relaxation time $\tau$. (b) The amplitudes of the anomalous and conventional parts versus the magnetic field. The parameters are $E_F=0.4$ eV, $v=0.3$ eVnm, $\alpha=0.2$ eVnm$^3$, $\beta=0.1$ eVnm, $r=0.1$, $\Delta=0.01$ eV in (a) and $\tau=0.1$ ps in (b).}\label{Fig:fig3}
\end{figure}

{\color{blue}\emph{Enhance anomalous planar Hall effect.}} -- 
{The missing anomalous planar Hall conductivity could be found if the third-order skew-scattering is larger.} This is because the conventional one $\sigma_{\text{H}}^\text{c}
\sim \sigma_\text{H}^\text{Boltz}$ is almost independent of $r$, while the third-order skew anomalous one is proportional to $r^{1.5}$. As revealed in Fig. \ref{Fig:fig2}(c), the amplitude of the anomalous part exceeds the conventional part when $r\approx100$. Thus the period of the total Hall conductivity shows a clear transition from $\pi$ to $2\pi/3$ with increasing the $r$. The anomalous planar Hall part arising from the third-order skew scattering dominates in the total conductivity at large ratio $r$.
Since third-order skew scattering is a higher-order term {with respect to the first line of the Feynman diagrams (Fig. 2)}, the ratio $r$ in a realistic sample is typically small. In order to observe the anomalous planar Hall effect in experiments, we propose to use strain to enhance it. The Zeeman energy $\Delta\sim0.003$ eV at $B=10$ T if $g=10$. 
Hence, because of its $\Delta^3$-behavior, the anomalous planar Hall effect is much smaller than conventional planar Hall effect. It is known that, the strain in the $x$-$y$ plane could induce a term $\beta k_x\sigma_z$ \cite{zhang2019prb,Zyuzin2020prb} that breaks the mirror symmetry and reduces the $C_{3v}$ symmetry to $C_{1v}$, which will result in another anomalous contribution with a $2\pi$ period in the planar Hall conductivity. Hence, the total planar Hall conductivity can be written as $\sigma_{\text{H}}=\sigma_{\text{H}}^\text{c}+\sigma_{\text{H}}^\text{A1}+\sigma_{\text{H}}^\text{A2}=C\sin2\theta+A_1\sin\theta+A_2\sin3\theta$ with $\sigma_{\text{H}}^\text{A1}$ and $\sigma_{\text{H}}^\text{A2}$ being the anomalous planar Hall conductivities induced by the strain and three-fold warping, respectively. 
As $\sigma_{\text{H}}^\text{A1}$ linearly depends on the magnetic field, it may be comparable to $\sigma_{\text{H}}^\text{c}$, especially in low-mobility samples. Therefore, the period may change from $\pi$ to $2\pi$ by decreasing $\tau$ as demonstrated in Fig. \ref{Fig:fig3}(a), where the $2\pi/3$ period still does not appear. We can extract their amplitudes from the conductivities at $\theta=\pi/4$, $\pi/2$, and $5\pi/4$ \cite{Supp-Weyl}, respectively, as shown in Fig. \ref{Fig:fig3}(b), where $\sigma_{\text{H}}^\text{A1}$ is out-of-phase relative to the $\sigma_{\text{H}}^\text{c}$ and $\sigma_{\text{H}}^\text{A2}$, and then $\sigma_{\text{H}}^\text{A1}\propto\Delta$, $\sigma_{\text{H}}^\text{A2}\propto\Delta^3$, and $\sigma_{\text{H}}^\text{c}\propto\Delta^2$, as expected. 

The X and $\Psi$ crossing diagrams (last three diagrams in Fig. \ref{Fig:surface}) are well-known to play significant roles in the anomalous Hall effect \cite{Ado2015epl,Ado2016prl}. We also consider their effects on the planar Hall effect and find that they lead to an anomalous planar Hall contribution \cite{Supp-Weyl}, but with a $\pi$ period instead of a $2\pi/3$ period. Nevertheless, its amplitude is three orders of magnitude smaller than that of the conventional planar Hall contribution. Therefore, our proposed data analysis scheme remains valid to distinguish the conventional and non-crossing anomalous planar Hall contributions, as long as the disorder scattering is weak. Furthermore, this scheme can distinguish the anomalous non-crossing and crossing contributions if the disorder scattering is strong.

{\color{blue}\emph{Acknowledgments}} --
We would like to thank Fengqi Song, Fucong Fei, Yoichi Ando, and Alexey Taskin for useful discussions. This work was supported by the National Key R\&D Program of China (2022YFA1403700), the National Natural Science Foundation of China (11534001, 11974249, and 11925402), the Innovation Program for Quantum Science and Technology (No. 2021ZD0302400), Guangdong province (2016ZT06D348, 2020KCXTD001), the Natural Science Foundation of Shanghai (19ZR1437300), and the Science, Technology and Innovation Commission of Shenzhen Municipality (ZDSYS20170303165926217, JCYJ20170412152620376, KYTDPT20181011104202253). The numerical calculations were supported by Center for Computational Science and Engineering of SUSTech.

%

\end{document}